\documentclass[prd,aps
,reprint,
,tightenlines,
superscriptaddress,showpacs
,nofootinbib
,eqsecnum,amsfonts,amsmath,floatfix
]{revtex4-1}
\usepackage{latexsym}
\usepackage{amssymb}
\usepackage{amsfonts}
\usepackage{amsmath}
\usepackage{bm}
\usepackage[dvips]{graphicx}
\usepackage[usenames]{color}
\usepackage[normalem]{ulem}
\usepackage{multirow}
\usepackage{natbib}
\usepackage{subfigure}
\allowdisplaybreaks

\newcommand{\simgt}{\lower.5ex\hbox{$\; \buildrel > \over \sim \;$}}
\newcommand{\simlt}{\lower.5ex\hbox{$\; \buildrel < \over \sim \;$}}

\newcommand{\F}{\mathcal{F}}
\begin{document}

\title{Comparison of post-Newtonian templates for extreme mass ratio inspirals}

  \author{Vijay~\surname{Varma}}
  \affiliation{Birla Institute of Technology and Science, Pilani 333031, India}
  \author{Ryuichi \surname{Fujita}}
  \affiliation{Departament de F\'isica, Universitat de les Illes Balears, 
Cra.\ Valldemossa Km.\ 7.5, Palma de Mallorca, E-07122 Spain}
  \author{Ashok~\surname{Choudhary}}
  \affiliation{Indian Institute of Science Education and Research, Pune 411008, India}
  \author{Bala R.~\surname{Iyer}}
  \affiliation{Raman Research Institute, Bangalore 560 080, India}

\begin{abstract}
Extreme mass ratio inspirals (EMRIs), the inspirals of  compact objects into  supermassive black holes, are  important gravitational wave sources for the Laser Interferometer Space Antenna (LISA). We study the performance of various post-Newtonian (PN) template families relative to the waveforms that are high-precision numerical solutions of the Teukolsky equation in the context of EMRI parameter estimation with LISA. Expressions for the time-domain waveforms TaylorT1, TaylorT2, TaylorT3, TaylorT4 and TaylorEt 
are derived up to 22PN order, i.e $\mathcal{O}(v^{44})$ ($v$ is the characteristic velocity of the binary) beyond the Newtonian term, for a test particle in a circular orbit around a Schwarzschild black hole. The phase difference between the above 22PN waveform families and numerical waveforms are evaluated during two-year inspirals for two prototypical EMRI systems with 
mass ratios $10^{-4}$ 
and 
$10^{-5}$. 
We find that the dephases (in radians) for TaylorT1 and TaylorT2, respectively, are about $10^{-9}$ ($10^{-2}$) and $10^{-9}$ ($10^{-3}$) for 
mass ratio $10^{-4}$ ($10^{-5}$). 
This suggests that using 22PN TaylorT1 or TaylorT2 waveforms for parameter estimation of EMRIs will result in accuracies comparable to numerical waveform accuracy for most of the LISA parameter space. 
On the other hand, from the dephase results, we find that TaylorT3, TaylorT4 and TaylorEt fare relatively poorly as one approaches the last stable orbit. 
 This implies that, as for comparable mass binaries using the 3.5PN phase of waveforms,
the 22PN TaylorT3 and TaylorEt approximants do not perform well enough for the EMRIs. 
The reason  underlying the  poor performance of TaylorT3, TaylorT4 and TaylorEt relative to TaylorT1 and TaylorT2 is finally examined. 

\end{abstract}

\date{\today}
  \pacs{
    04.25.Nx   
  }
\maketitle
\section{Introduction}
The inspiral of a stellar-mass compact object into a supermassive black hole (SMBH) is one of the most promising gravitational wave (GW) sources for space-based detectors such as eLISA.\footnote{evolved Laser Interferometer Space Antenna, also known as NGO (New Gravitational-Wave Observatory).} The compact object typically has a mass of the order of a few solar masses while the SMBHs detectable by eLISA are in the mass range $10^5M_{\odot}-10^7M_{\odot}$. As the mass ratio for these binaries is typically around $10^5$, these systems are called extreme mass ratio inspirals (EMRIs). Gravitational waves from EMRIs can provide information about the parameters of the central black hole such as its spin, mass and details of its stellar surroundings while also facilitating strong field tests of general relativity (GR) \cite{sfld1,sfld2,sfld3}.

However, the gravitational wave signal is buried in a background of noise and the signal needs to be extracted using data analysis techniques such as matched filtering. 
Matched filtering requires accurate templates of the gravitational waveform.
For eLISA EMRI parameter estimation, the GW phase errors of the template with respect to the true signal should be less than $10$ milliradians \cite{accu}. Considering eLISA is expected to detect 10-1000 EMRIs during its mission \citep{eventrate1,eventrate2,eventrate3}, the search for accurate waveforms for these systems is justified.

Post-Newtonian (PN) theory provides a method to predict the gravitational waveform for the early phase of inspiraling compact binaries~\cite{Blanchet:2006zz}. 
However, since the PN approximation breaks down near the last stable orbit (LSO), 
numerical relativity (NR) 
waveforms are required beyond this point. Nevertheless, PN waveforms in the early inspiral phase are required to calibrate 
the NR waveforms because the computational cost for NR is very high. 
PN waveforms can be matched with the NR waveforms in the late inspiral and the subsequent merger and ringdown phases \cite{a1,bpjc} to provide a cheaper alternative to using NR for the complete inspiral. 
PN waveforms for nonspinning comparable mass binaries in quasicircular orbits are known up to 3.5PN order beyond the Newtonian term \cite{biops,dis1,dis2}. Within the post-Newtonian formalism, several nonequivalent template families such as TaylorT1, TaylorT2, TaylorT3, TaylorT4, TaylorEt and TaylorF2, among others, are possible. These 3.5PN template families were discussed extensively in Ref. \cite{biops}. 
 It is found that for $M<12\,M_\odot$,where $M$ is the total mass of the binary,  these 3.5PN template families 
except for TaylorT3 and TaylorEt are equally good for the detection of 
gravitational waves using ground-based detectors.

The mass ratio for EMRIs is very small, $10^{-4}-10^{-7}$, and one can apply 
black hole perturbation theory to compute  the gravitational wave emission
 using the mass ratio as an expansion parameter~\cite{Sasaki:2003xr}. 
Using black hole perturbation theory, 
one can go to a much  higher order of PN iteration for gravitational waves 
than for comparable masses using  standard PN theory. 
Recently, 22PN waveforms have been calculated for a test particle in a circular orbit around a Schwarzschild black hole~\cite{f22} by solving the Teukolsky equation~\cite{Teukolsky1973}, which is a fundamental equation of the black hole perturbation theory. 
It is shown that the 22PN gravitational waveforms achieve data analysis accuracies comparable to waveforms resulting from high-precision numerical solutions of the Teukolsky equation.\footnote{The Teukolsky equation is a first-order perturbation equation, in which the particle moves on geodesics of the black hole. Over time scales of the inverse of the mass ratio, the orbit deviates from the geodesic because of the gravitational self-force~\cite{Poisson:2011nh,Barack:2009ux}. Using numerical results for the full relativistic first-order gravitational self-force in Ref.~\cite{Barack:2007tm}, the dephase due to the gravitational self-force is estimated as a few radians~\cite{Burko:2013cca,Warburton:2011fk}. Thus, the gravitational self-force should be taken into account in the future.}
In this paper we extend this study by calculating the different template families mentioned above up to 22PN order using 
the 22PN energy flux derived in Ref.~\cite{f22}. 
We then investigate the performance of these Taylor approximants by evaluating the phase difference between these approximants and the waveforms that are high-precision numerical solutions  of the Teukolsky equation 
in Refs.~\cite{fnum1,fnum2} 
over a two-year inspiral for two systems, one in the early inspiral phase and the other in the late inspiral phase of the eLISA frequency band. We find that TaylorT1 (which was also investigated in Ref.~\cite{f22}) and TaylorT2 provide the best matches to numerical waveforms while the phase difference increases by a few orders of magnitude for TaylorT3, TaylorT4 and TaylorEt. 
 These investigations extend the results for comparable mass binaries 
in Ref.~\cite{biops} that TaylorT3 and TaylorEt approximants are considerably different 
from the others and perform relatively poorly. 
We also discuss why the performance of TaylorT3, TaylorT4 and TaylorEt 
in the test particle limit becomes worse than the others. 
This may provide insights which should be kept in mind when one constructs 
new PN template families.

This paper is organized as follows. In Sec. II we discuss the various template families along with the relevant initial conditions and calculate these approximants upto 22PN order. In Sec. III we evaluate the dephase between these different PN waveform approximants and a fiducial waveform that is a 
high-precision numerical  solution  of the Teukolsky equation 
during two-year inspirals. In Sec. IV we summarize our main conclusions. Since the 22PN Taylor approximants are too large to be shown in this paper, we only show them up to 4.5PN order. The 22PN expressions for the approximants will be publicly available online~\cite{sm}. Throughout this paper we use units $c=G=1$.

\section{The post-Newtonian approximants}
\label{secII}
Post-Newtonian approximation treats the early stages of adiabatic inspiral of compact binaries as a perturbative model and expresses a binding energy, $E(v)$ and a flux, $\F(v)$ associated with the gravitational wave as a power series in $v$, where $v=(\pi M F)^{1/3}$ is the characteristic velocity, $M$ is the total mass  and $F$ is the gravitational wave frequency of the binary. Here, adiabatic inspiral implies that the inspiral time scale is much larger than the orbital time scale. For restricted waveforms,\footnote{Restricted waveforms are obtained by retaining only the leading harmonic in the GW signal. For these waveforms the gravitational wave phase is twice the orbital phase.} under the adiabatic approximation, the standard energy balance equation, $dE_{\rm tot}/dt=-\F$ gives us the following pair of coupled differential equations for the orbital phasing formula \cite{biops,dis1,dis2}:

\begin{subequations}
\label{phasing}
\begin{equation}
\frac{d\phi}{dt}-\frac{v^3}{M}=0,
\label{phasinga}
\end{equation}
\begin{equation}
\frac{dv}{dt}+\frac{\F(v)}{M E'(v)}=0.
\label{phasingb}
\end{equation}
\end{subequations}

Here, $E'(v)$ is the derivative of the binding energy with respect to the characteristic velocity, $v$. The binding energy, $E$ is related to the total energy, $E_{\rm tot}$ by $E_{\rm tot}=M(1+E)$.

The phasing formula can also be expressed in the following equivalent parametric form, where $t_{\rm ref}$ and $\phi_{\rm ref}$ are integration constants and $v_{\rm ref}$ is an arbitrary reference velocity:
\begin{subequations}
\label{parametric}
\begin{equation}
t(v)=t_{\rm ref} + M\int_v^{v_{\rm ref}} \frac{E'(v)}{\F(v)}\,dv,
\label{parametrica}
\end{equation}
\begin{equation}
\phi(v)=\phi_{\rm ref} + \int_v^{v_{\rm ref}}v^3\frac{E'(v)}{\F(v)}\,dv.
\label{parametricb}
\end{equation}
\end{subequations}

Recently, the 22PN order energy flux for EMRIs has been calculated \cite{f22}. 
For the extreme mass ratio binaries, $E(v)$ is known exactly, 
see e.g. Ref.~\cite{Bardeen:1972fi}, 
and can be expanded and truncated to any required PN order.\footnote{In contrast, for comparable mass binaries, $E(v)$ cannot be derived exactly and
is computed as one of the conserved quantities associated with a specified-order
 PN iteration of the equation of motion. Currently it is known up to 3PN 
for  nonspinning binaries in general orbits, see e.g. Ref.~\cite{Blanchet:2006zz}. The flux function $F(v)$ on the other hand is
known only as a PN expansion in both the test particle and comparable mass
cases albeit to a much higher PN order in the test particle case (22PN)
relative to the comparable mass case (3.5PN).}
We present for the convenience of the reader, expressions for 4PN $E(v)$ and 4.5PN $\F(v)$ that are inputs needed to derive  the 4.5PN results displayed explicitly in later sections for brevity of presentation.
In these expressions, $m_1$ and $m_2$ are the masses of the test particle and the SMBH,  
respectively, $ M=m_1+m_2$ is the total mass, 
$\nu=m_1m_2/M^2$ is the symmetric mass ratio and $\gamma=0.577216...$ is the Euler constant.

\begin{widetext}
\begin{equation}
E_4(v)=-\frac{1}{2}\nu v^2\left[1-\frac{3 v^2}{4}-\frac{27 v^4}{8}-\frac{675 v^6}{64}-\frac{3969
   v^8}{128}\right],
\label{energy}
\end{equation}

\begin{eqnarray}
\label{flux}
\F_{4.5}(v)=&& \frac{32}{5}\nu^2 v^{10} \biggl[1-\frac{1247 v^2}{336}+4 \pi  v^3-\frac{44711 v^4}{9072}-\frac{8191 \pi  v^5}{672} +v^6
   \biggl\{\frac{6643739519}{69854400}+\frac{16 \pi ^2}{3}-\frac{1712 \gamma
   }{105}  -\frac{856 }{105} \log (16 v^2)\biggr\} \nonumber\\&& -\frac{16285 \pi 
   v^7}{504} +v^8 \biggl\{-\frac{323105549467}{3178375200}-\frac{1369 \pi ^2}{126}+\frac{232597 \gamma
   }{4410}+\frac{39931 }{294} \log(2)-\frac{47385 }{1568} \log (3) +\frac{232597 }{4410} \log (v)\biggr\} \nonumber\\&& +\pi v^9 \biggl\{\frac{265978667519}{745113600}-\frac{6848 \gamma}{105}-\frac{3424}{105} \log (16 v^2) \biggr\} \biggr].
\end{eqnarray}
\end{widetext}

Approximate waveforms are obtained by inserting the expressions for $E(v)$ and $\F(v)$ at consistent PN order into the phasing formula - these waveforms are referred to as Taylor approximants. There are several ways of inserting these expressions into the phasing formula leading to different approximants such as TaylorT1, TaylorT4, TaylorT2, TaylorT3, TaylorEt and TaylorF2 \cite{biops}. We have calculated these approximants up to 22PN order for the EMRI case. We shall now discuss these approximants while presenting our results up to 4.5PN. The method for calculating the phase of the gravitational waveform is left for the next section. The complete 22PN expressions will be available online~\cite{sm}.

\subsection{TaylorT1}
\label{Taylort1}
The TaylorT1 approximant is obtained by using the expressions for binding energy, $E(v)$, and flux, $\F(v)$, as they appear in Eqs.~(\ref{energy}) and (\ref{flux}) in the phasing formula, Eq.~(\ref{phasing}), and solving the resulting equations involving the rational polynomial  $\F(v)/E'(v)$ numerically:
\begin{subequations}
\label{T1}
\begin{equation}
\frac{d\phi^{(\rm T1)}}{dt}-\frac{v^3}{M}=0,
\label{T1a}
\end{equation}
\begin{equation}
\frac{dv}{dt}+\frac{\F(v)}{M E'(v)}=0.
\label{T1b}
\end{equation}
\end{subequations}

In the above equations $v \equiv v^{(\rm T1)}$, but for simplicity we omit the superscript.
The expressions for $E(v)$ and $\F(v)$ are to be truncated to consistent PN order to obtain the approximant of that order. This is followed for all the approximants in this section.

\subsection{TaylorT4}
\label{TaylorT4}
TaylorT4 goes one step further than TaylorT1 by expanding the rational polynomial $\F(v)/E'(v)$ and truncating it to the required PN order~\cite{Boyle:2007ft}. The characteristic velocity, $v^{(\rm T4)}(t)\equiv v(t)$ at 4.5PN is given for TaylorT4 by
\begin{widetext}
\begin{eqnarray}
\label{T4}
\frac{dv}{dt}=&& \frac{32}{5} \frac{\nu}{M} v^{9} \biggl[1-\frac{743 v^2}{336}+4 \pi  v^3+\frac{34103 v^4}{18144}-\frac{4159 \pi  v^5}{672} +v^6\biggl\{\frac{16447322263}{139708800}+\frac{16 \pi ^2}{3}-\frac{1712 \gamma
   }{105}-\frac{856 }{105} \log (16 v^2) \biggr\} \nonumber\\&& -\frac{4415 \pi 
   v^7}{4032} +v^8 \biggl\{\frac{3959271176713}{25427001600}-\frac{361 \pi ^2}{126}+\frac{124741 \gamma
   }{4410}+\frac{127751}{1470}  \log(2)-\frac{47385 }{1568} \log (3)+\frac{124741 }{4410} \log (v)\biggr\} \nonumber\\&& +\pi v^9 \biggl\{\frac{343801320119}{745113600}-\frac{6848 \gamma
   }{105}-\frac{3424 }{105} \log (16 v^2) \biggr\} \biggr].
\end{eqnarray}
Similarly to TaylorT1, Eq.~(\ref{phasinga}) gives the evolution of the orbital phase for TaylorT4.

\subsection{TaylorT2}
\label{TaylorT2}
TaylorT2 uses the parametric form of the phasing formula, Eq.~(\ref{parametric}). The ratio $E'(v)/\F(v)$ is expanded and truncated to the required PN order. Upon integration we obtain the following equations for $\phi(v)$ and $t(v)$ at 4.5PN order:
\begin{subequations}
\label{T2}
\begin{eqnarray}
\phi^{(\rm T2)}_{4.5}(v)=&&\phi^{(\rm T2)}_{\rm ref} -\frac{1}{32 \nu v^5} \biggl[1+\frac{3715 v^2}{1008}-10 \pi 
   v^3+\frac{15293365
   v^4}{1016064}+\frac{38645 \pi}{672} v^5 
   \log \left(\frac{v}{v_{\rm lso}}\right) +v^6 \biggl\{\frac{12348611926451}{18776862720}-\frac{160 \pi^2}{3} \nonumber\\&& -\frac{1712
   \gamma}{21}   -\frac{856}{21}\log(16 v^2)\biggr\} +\frac{77096675 \pi 
   v^7}{2032128} +v^8 \biggl\{\frac{2550713843998885153}{221446
   8081745920}-\frac{45245 \pi^2}{756}-\frac{9203 \gamma}{126}  \nonumber\\&& -\frac{252755 }{2646} \log(2) -\frac{78975}{1568} \log
   (3)-\frac{9203}{126} \log(v)\biggr\}  \nonumber\\&& +\pi v^9 \biggl\{-\frac{93098188434443}{150214901760} +\frac{80 \pi^2}{3} +\frac{1712 \gamma }{21} +\frac{856}{21} \log(16 v^2) \biggr\} \biggr],
\label{T2a}
\end{eqnarray}
\begin{eqnarray}
t^{(\rm T2)}_{4.5}(v)=&& t^{(\rm T2)}_{\rm ref} -\frac{5 M}{256 \nu v^8}\biggl[ 1+\frac{743 v^2}{252}-\frac{32 \pi 
   v^3}{5}+\frac{3058673
   v^4}{508032}-\frac{7729 \pi 
   v^5}{252} +v^6 \biggl\{-\frac{10052469856691}{23471078400} +\frac{128 \pi
   ^2}{3}  +\frac{6848 \gamma}{105}  \nonumber\\&& +\frac{3424 }{105} \log(16 v^2) \biggr\} -\frac{15419335 \pi 
   v^7}{127008} +v^8 \biggl\{\biggl(\frac{2496799162103891233}{461347517030400} -\frac{18098 \pi^2}{63} -\frac{36812 \gamma}{105} -\frac{202204 }{441} \log(2) \nonumber\\&& -\frac{47385 }{196} \log(3)\biggr) \log (v)-\frac{18406}{105} \log^2(v)\biggr\} \nonumber\\&& +\pi v^9 \biggl\{ -\frac{102282756713483}{23471078400}  +\frac{512 \pi^2}{3} +\frac{54784 \gamma}{105}+\frac{27392 }{105} \log(16 v^2) \biggr\} \biggr].
\label{T2b}
\end{eqnarray}
\end{subequations}
Here $t_{\rm ref}$ and $\phi_{\rm ref}$ are integration constants. $t_{\rm ref}$ is fixed by setting $t=0$ when $v=v_0$, the initial velocity.

\subsection{TaylorT3}
\label{TaylorT3}
To get the TaylorT3 approximant, the expression for $t(v)$ generated in TaylorT2 is inverted to get $v(t)$. This is then used to obtain $\phi(t) \equiv \phi(v(t))$. The TaylorT3 also gives the instantaneous gravitational wave frequency $F$ by $F\equiv d\phi/(\pi dt)= v^3/(\pi M)$. TaylorT3 approximant at 4.5PN order is given by 
\begin{subequations}
\label{T3}
\begin{eqnarray}
\phi^{(\rm T3)}_{4.5}(t)=&&\phi^{(\rm T3)}_{\rm ref} -\frac{1}{\nu \theta^5} \biggl[ 1 + \frac {3715 \theta^2} {8064} - \frac {3 \pi
         \theta^3} {4} + \frac {9275495\theta^4} {14450688} + \frac{38645 \pi  \theta^5 }{21504} \log\left(\frac{\theta}{\theta_{\rm lso}}\right) +\theta^6 \biggl\{\frac {831032450749357} {57682522275840} - \frac {53 \pi^2} {40} \nonumber\\&& - \frac {107 \gamma} {56}  -\frac {107 } {56} \log(2\theta) \biggr\} + \frac {188516689 \pi 
        \theta^7} {173408256} + \theta^8 \biggl\{\frac {11715802333726918585} {2073248288647151616}- \frac {191257 \pi^2} {387072} - \frac {312247 } {451584} \log^2 (2) \nonumber\\&& + \frac {2446934992845948193 } {188967942975651840} \log(2)    - \frac {236925 } {401408} \log (2)\log (3) - \frac {78975} {401408}  \log(3)  - \frac {\gamma (208343 + 386526 \log(2))} {451584} \nonumber\\&& - \frac {45245 \pi ^2} {64512}  \log(2)  + \biggl(-\frac {2583981498376602913} {188967942975651840}  + \frac {45245 \pi^2} {64512}+ \frac {9203 \gamma } {10752} + \frac {14873 } {56448} \log(2) + \frac {236925 } {401408} \log(3)\biggr) \log(\theta) \nonumber\\&& +\frac {9203 \log^2 (\theta)} {21504} \biggr\}   + \pi \theta^9 \biggl\{ \frac {587519428177201} {192275074252800} - \frac {33 \pi^2} {800}   - \frac {321 \gamma } {1120} -\frac {321 } {1120} \log(2\theta)\biggr\} \biggr],
\label{T3a}
\end{eqnarray}
\begin{eqnarray}
F^{(\rm T3)}_{4.5}(t)=&& \frac{\theta^3}{8 \pi M} \biggl[ 1 + \frac {743 \theta^2} {2688} - \frac {3 \pi 
        \theta^3} {10} + \frac {1855099
        \theta^4} {14450688} - \frac {7729 \pi 
        \theta^5} {21504}  + \theta^6
   \biggl\{  - \frac {720817631400877} {2884126113
          79200}  + \frac {53 \pi
           ^2} {200} + \frac {107 \gamma
         } {280}  +\frac {107 } {280} \log(2\theta) \biggr\} \nonumber\\&& - \frac {188516689 \pi 
        \theta^7} {433520640}  + \theta^8
   \biggl\{ - \frac {2033421792006076349} {3
          101012397549158400} + \frac {33589 \pi
           ^2} {215040}  + \frac {312247 } {752640} \log^2(2)  - \frac {2463531507726173473} {314946571626086400} \log(2) \nonumber\\&& + \frac {142155 } {401408} \log (2)
           \log (3)+ \frac {\gamma 
          (79501 + 386526 \log
             (2))} {752640}   + \frac {9049 \pi ^2 } {21504} \log(2)   + \biggl (\frac {253006
              6816481608993} {314946571626086400} - \frac {9049 \pi^2} {21504} \nonumber\\&& - \frac
            {9203 \gamma } {17920}  - \frac {14873 } {94080} \log(2)   - \frac {142155 } {401408} \log(3)\biggr) \log
        (\theta) -\frac {9203 } {35840} \log^2 (\theta)\biggr\}  \nonumber\\&& + \pi \theta^9 \biggl\{ -\frac {573742575758641} {240343842816000} + \frac {33 \pi^2} {1000}  + \frac {321 \gamma
             } {1400}  +\frac {321} {1400}  \log(2\theta) \biggr\} \biggr],
\label{T3b}
\end{eqnarray}
\end{subequations}
where $\theta(t)$ is given by $\theta=[\nu(t_{\rm ref} - t)/(5M)]^{-1/8}$. Given an initial velocity $v_0$, one can find the initial frequency $F_0$, by $F_0=v_0^3/(\pi M)$. To find $t_{\rm ref}$, one solves Eq.~(\ref{T3b}) at $t=0$ and $F=F_0$.

\subsection{TaylorEt}
\label{TaylorEt}
TaylorEt is expressed as a power series of a new function, $\zeta=-2E/\nu$~\cite{Gopakumar:2007jz}. Equation (\ref{energy}) for $E(v)$ can be expressed in terms of $x=v^2$ to get $\zeta(x)$. This is then inverted to obtain $x(\zeta)$:

\begin{equation}
x(\zeta)= \zeta \biggl[1 + \frac {3 \zeta} {4} + \frac {9 \zeta^2} {2} + \frac {405 \
\zeta^3} {16} + \frac {2511\zeta^4} {16} \biggr].
\label{x}
\end{equation}

From the phasing formula Eq.~(\ref{phasinga})  and Eq.~(\ref{x}) we get an expression for the evolution of phase in terms of $\zeta$ [cf. Eq.~(\ref{Eta})]. Under the new variable, $\zeta$, Eq.~(\ref{phasingb}) of the phasing formula transforms to
\begin{equation}
\frac{d\zeta}{dt}=\frac{2\F(v(\zeta))}{\nu M}.
\label{phasingEt}
\end{equation}

The TaylorEt approximant is, essentially, the gravitational wave phasing equations expressed in terms of $\zeta$. At 4.5PN order, it is given by 
\begin{subequations}
\label{Et}
\begin{eqnarray}
\label{Eta}
\frac{d\phi^{(\rm Et)}(t)}{dt}=&& \frac{\zeta^{3/2}}{M} \biggl[ 1 + \frac {9 \zeta} {8} + \frac {891 \zeta^2} {128} + \frac {41445 \
\zeta^3} {1024} + \frac {8413875\zeta^4} {32768} \biggr],
\end{eqnarray}
\begin{eqnarray}
\label{Etb}
\frac{d\zeta}{dt}=&& \frac{64 \nu \zeta^{5}}{5M} \biggl[1 + \frac {13\zeta} {336} + 
 4\pi \zeta^{3/2} + \frac {117857\zeta^2} {18144}  + \frac {4913\pi} {672}\zeta^{5/2} +\zeta^3
   \biggl\{\frac {37999588601} {279417600} + \frac {16\pi^2} {3} - \frac {1712\gamma} {105} 
  - \frac {856} {105} \log (16\zeta) \biggr\} \nonumber\\&&  + \frac {129817\pi } {2304} \zeta^{7/2}  +\zeta^4
   \biggl\{ \frac {3677099151569} {5085400320} + \frac {2663\pi^2} {126} - \frac {198827\gamma} 
{4410}  - \frac {87961} {1470} \log (2) - 
\frac {47385} {1568} \log (3)  - \frac {198827}{8820} \log (\zeta) \biggr\} \nonumber\\&&    + \pi \zeta^{9/
      2}\biggl\{\frac {1130297606413} {1490227200} - \frac 
{6848\gamma} {105}  - \frac {3424} 
{105}\log (16 \zeta)\biggr\} \biggr].
\end{eqnarray}
\end{subequations}

As in the case of TaylorT3, we can find $F_0$, given $v_0$. Noting that $F\equiv d\phi/(\pi dt)$, initial conditions for TaylorEt can be set up by solving Eq.~(\ref{Eta}) for $\zeta_0$ by setting the left-hand side to $\pi F_0$.


\subsection{TaylorF2}
\label{TaylorF2}
TaylorF2 is a Fourier-domain approximant based on the stationary phase approximation (SPA). Under the SPA, the waveform in the Fourier domain is expressed as 
\begin{equation}
\label{spa}
\tilde{h}^{\rm spa}(f)=\frac{a(t_{f})}{\sqrt{\dot{F}(t_{f})}} e^{i[\psi_{f}(t_{f})-\pi/4]},  \psi_{f}(t) \equiv 2\pi ft-2\phi(t),
\end{equation}
where $t_f$ is the saddle point, defined by solving for $t$ when $d\psi_f(t)/dt=0$, i.e the time $t_f$ when the gravitational wave frequency $F(t)$ becomes equal to the Fourier variable, $f$. In the adiabatic approximation [where Eqs.~(\ref{parametrica}) and (\ref{parametricb}) hold], the values of $t_f$ and $\psi_f(t_f)$ are given by

\begin{subequations}
\label{F2par}
\begin{equation}
t_f=t_{\rm ref} + M\int_{v_f}^{v_{\rm ref}} \frac{E'(v)}{\F(v)}\,dv,
\label{F2para}
\end{equation}
\begin{equation}
\psi_f(t_f)=2\pi f t_{\rm ref} -\phi_{\rm ref}+2\int_{v_f}^{v_{\rm ref}}(v_f^3-v^3)\frac{E'(v)}{\F(v)}\,dv,
\label{F2parb}
\end{equation}
\end{subequations}
where $v_f=(\pi M f)^{1/3}$. 

Using expressions of energy and flux and expanding the ratio $E'(v)/\F(v)$ in 
Eq.~(\ref{F2par}) to consistent PN order leads to an expression which can be integrated explicitly resulting in the TaylorF2 approximant. The phase of the Fourier-domain waveform up to 4.5PN order is given by
\begin{eqnarray}
\label{F2}
\psi^{(\rm F2)}_{4.5}(f) =&& 2 \pi f t_{\rm c} -\phi_{\rm c}-\frac{\pi}{4} +\frac{3}{128 \nu v^5} \biggl[1 + \frac {3715 v^2} {756} - 16 \pi  v^3 + \frac {15293365
       v^4} {508032} + \frac {38645\pi v^5} {252}  \log \left(\frac{v}{v_{\rm lso}}\right) \nonumber\\&& + v^6
   \biggl\{\frac {11583231236531} {4694215680}  - \frac {640 \pi ^2} {3}  - \frac {6848
          \gamma } {21}  -\frac {3424 } {21} \log (16v^2) \biggr\}  + \frac {77096675 \pi  v^7} {254016}  + v^8
   \biggl\{  \biggl (-\frac {2550713843998885153} 
{276808510218240} \nonumber\\&& + \frac {90490 \pi ^2} {189} + \frac {36812 \gamma } {63}   + \frac 
{1011020 } {1323} \log(2)  + \frac {78975 } {196} \log (3) \biggr) \biggl( \log (v) - \frac{1}{3} \biggr) +\frac {18406} {63}  \log^2 (v) \biggr\}  \nonumber\\&& +\pi v^9
   \biggl\{  \frac {105344279473163} {18776862720} - \frac {640 \pi^2} {3} - \frac {13696\gamma} {21}   -\frac {6848} {21} \log (16 v^2) 
 \biggr\} \biggr],
\end{eqnarray}
\end{widetext}
where $t_c$ and $\phi_c$ can be chosen arbitrarily and $v=(\pi M f)^{1/3}$.

The behavior of TaylorF2 has already been investigated up to 3.5PN order for comparable mass binaries \cite{biops,dis1,dis2}. One must keep in mind that the stationary phase approximation, on which TaylorF2 is based, 
is valid only up to 4.5PN order \cite{dkpo,dis3}. Thus, beyond 4.5PN the Fourier transform of the waveform 
has correction terms for 
the stationary phase approximation to the Fourier transform in Eq.~(\ref{spa}). 
However, as a start, in this paper, we have obtained the TaylorF2 approximant 
up to 22PN order by assuming Eq.~(\ref{spa}) is valid even beyond 4.5PN. 
Further studies, computing terms beyond the leading \cite{biops,dis1,dis2} 
are needed to look for a good frequency-domain approximant
at higher PN orders and will be investigated in the future.

\section{Comparison with high-precision numerical solutions of the Teukolsky equation}
\label{secIII}
To investigate the behavior of different analytical PN families described in Sec. \ref{secII} we calculate the phase of the gravitational wave signal during a two-year quasicircular inspiral of two systems of binaries called System-I and System-II as considered in Refs.~\cite{f22,f14,fi,ybhmp}. We compare this phase with the phase calculated using waveforms that are 
high-precision numerical solutions  of the Teukolsky equation, 
the difference between these phases is called the dephase. System-I has masses $(m_1, m_2)=(10,10^5)M_{\odot}$ with $m_1/m_2=10^{-4}$; it inspirals from $r_{\rm in} \simeq 29M$ to $r_{\rm fin} \simeq 16M$ during a two-year period, with gravitational wave frequencies in the range $f_{\rm GW}\in[4\times10^{-3},10^{-2}]$Hz. System-II has masses $(m_1, m_2)=(10,10^6)M_{\odot}$ with $m_1/m_2=10^{-5}$; it inspirals from $r_{\rm in} \simeq 11M$ to $r_{\rm fin} \simeq 6M$ (LSO) during a two-year period, with gravitational wave frequencies in the range $f_{\rm GW}\in[1.8\times10^{-3},4.4\times10^{-3}]$Hz. In the frequency band of eLISA, System-I corresponds to the early inspiral phase of an EMRI, while System-II corresponds to the late inspiral phase. Note that the phases shown in the figures of this section are gravitational wave phases which are twice the orbital phases.

Matched filtering can give signal-to-noise ratios (SNRs) of up to $\rho \sim 100$ \cite{lisasnr1,lisasnr2} for the strongest EMRI signals detectable by eLISA. This means eLISA can detect phases up to an accuracy of order $1/\rho \sim 10$ milliradians \cite{accu}. Therefore, while considering the dephase results of this paper, we expect PN waveforms to have accuracies comparable to those provided by numerical waveforms if the dephase is less than $10^{-2}$ radians.

The numerical waveforms we use are based on those in Refs. \cite{fnum1,fnum2}, 
which solve the Teukolsky equation. 
The truncation of the $l$ mode limits the accuracy of the numerical calculations. We use the same data generated for Ref. \cite{f22}, which is based on $l=25$ calculations and gives relative error better than $10^{-14}$ up to the LSO.

\subsection{Dephase between TaylorT1 and numerical results}
\label{DephaseT1}
\begin{figure*}[htb]
\begin{center}

\subfigure[System-I for TaylorT1\label{fig1a}]{\includegraphics[scale=0.75]{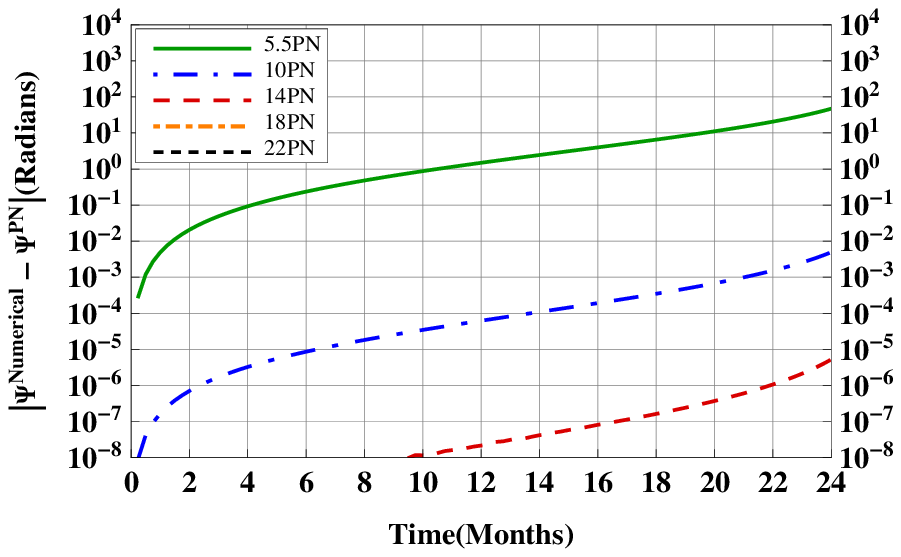}}\quad
\subfigure[System-II for TaylorT1\label{fig1b}]{\includegraphics[scale=0.75]{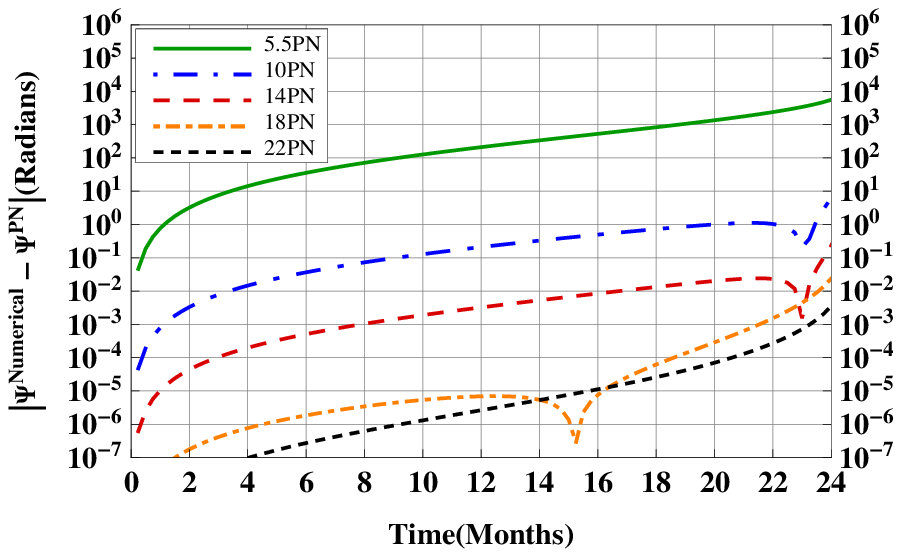}}\quad
\caption{Absolute value of the dephase between TaylorT1 PN waveforms and numerical waveforms during two-year inspirals as a function of time in months. The left panel shows the dephase for System-I having masses $(m_1, m_2)=(10,10^5)M_{\odot}$ for $r_{\rm in} \simeq 29M$ to $r_{\rm fin} \simeq 16M$ sweeping GW frequencies in the range $f_{\rm GW}\in[4\times10^{-3},10^{-2}]$Hz. The right panel shows the dephase for System-II having masses $(m_1, m_2)=(10,10^6)M_{\odot}$ for $r_{\rm in} \simeq 11M$ to $r_{\rm fin} \simeq 6M$ (LSO) sweeping GW frequencies in the range $f_{\rm GW}\in[1.8\times10^{-3},4.4\times10^{-3}]$Hz. System-I (System-II) corresponds to the early (late) inspiral phase of the eLISA frequency band. Note that the dephase between 18PN (22PN) TaylorT1 waveforms and numerical waveforms at the end of the two-year inspiral for System-I is about $8 \times 10^{-9}$ ($10^{-9}$) radians, which falls below the lowest value of dephase in the left panel.
}\label{fig1}
\end{center}
\end{figure*}

The dephase between TaylorT1 and numerical waveforms (cf. Fig. \ref{fig1}) was shown in Ref. \cite{f22}. We present the results here for comparison.\footnote{Fig. \ref{fig1} is slightly different from the dephase results of Ref. \cite{f22}. This is because in Ref. \cite{f22} the TaylorT1 phase was calculated without expanding $dE/dv$ in $v$.} For System-I (II), the absolute values of the dephasing between the TaylorT1 waveforms and the numerical waveforms after the two-year inspiral are about $7\times10^1 (8\times10^3), 7\times10^{-3} (8), 7\times10^{-6} (5\times10^{-1}), 8\times10^{-9} (3\times10^{-2})$ and $10^{-9} (5\times10^{-3})$ radians for 5.5PN, 10PN, 14PN, 18PN and 22PN respectively. It is also suggested in Ref. \cite{f22} that using 22PN TaylorT1 waveforms for EMRIs will result in accuracy of data analysis comparable to those resulting from high-precision numerical waveforms as the dephase is less than $10^{-2}$ radians for most of the parameter space of eLISA. We also note that 10PN TaylorT1 waveforms may be 
comparable in accuracy of data analysis to numerical waveforms for System-I.

We now extend this study by investigating the behavior of other PN Taylor families by evaluating their dephases during the same inspiral. 

\subsection{Dephase between TaylorT4 and numerical results}
\label{DephaseT4}
The calculation of phase of TaylorT4 is very similar to that of TaylorT1 and numerical waveforms. We use the relation
\begin{equation}
\phi(t)=\int_{0}^{t} (d\phi/dt') \,dt' =\int_{v_0}^{v(t)} 
\frac{(d\phi/dt')}{(dv'/dt')} \,dv'.  
\end{equation}
Here, $d\phi/dt$ is given by Eq.~(\ref{phasinga}), $v_0$ 
is the velocity at the starting 
of the two-year inspiral and $v(t)$ can be obtained by solving the equation 
$t-\int_{v_0}^{v(t)} 1/(dv/dt) \,dv=0$ for a given time $t$. 
For TaylorT4 $dv/dt$ is given by Eq.~(\ref{T4}) while for numerical waveforms $dv/dt$ is obtained from the solution of Teukolsky equation \cite{fnum1,fnum2,f22}.

\begin{figure*}[htb]
\begin{center}
 
    \subfigure[System-I for TaylorT4\label{fig2a}]{\includegraphics[scale=0.85]{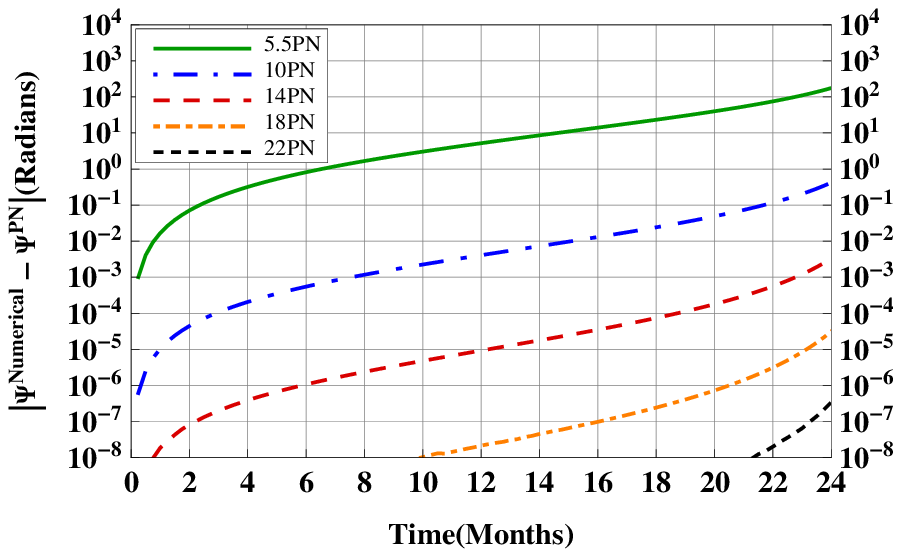}}\quad
    \subfigure[System-II for TaylorT4\label{fig2b}]{\includegraphics[scale=0.85]{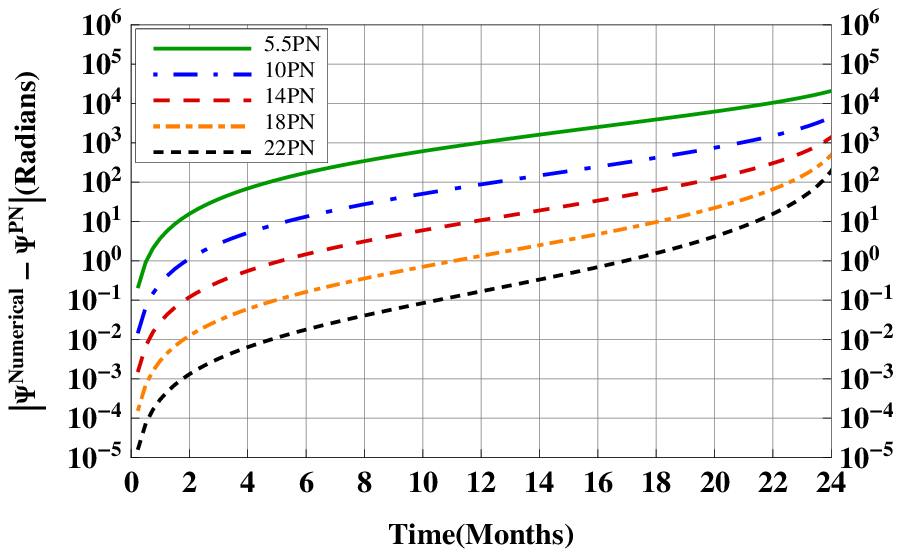}}\quad
    \subfigure[System-I for TaylorT2\label{fig2c}]{\includegraphics[scale=0.85]{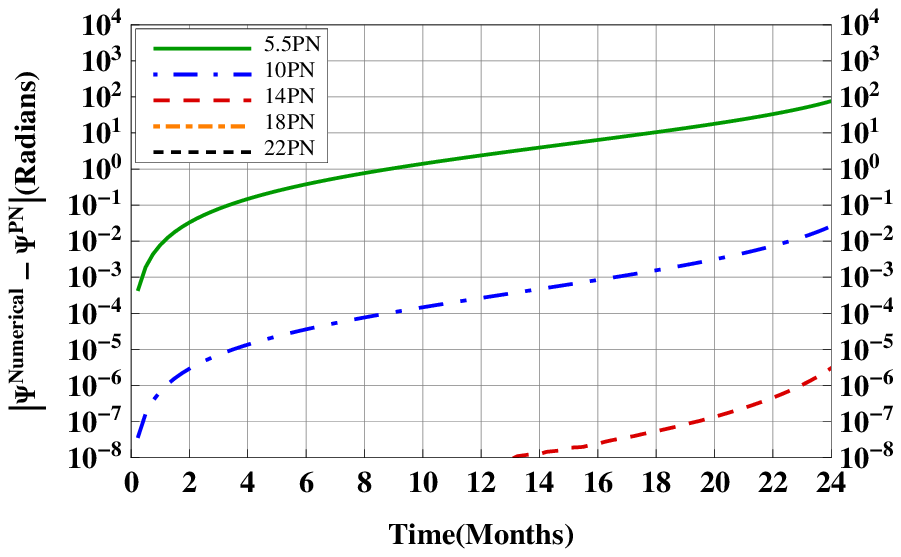}}\quad
		\subfigure[System-II for TaylorT2\label{fig2d}]{\includegraphics[scale=0.85]{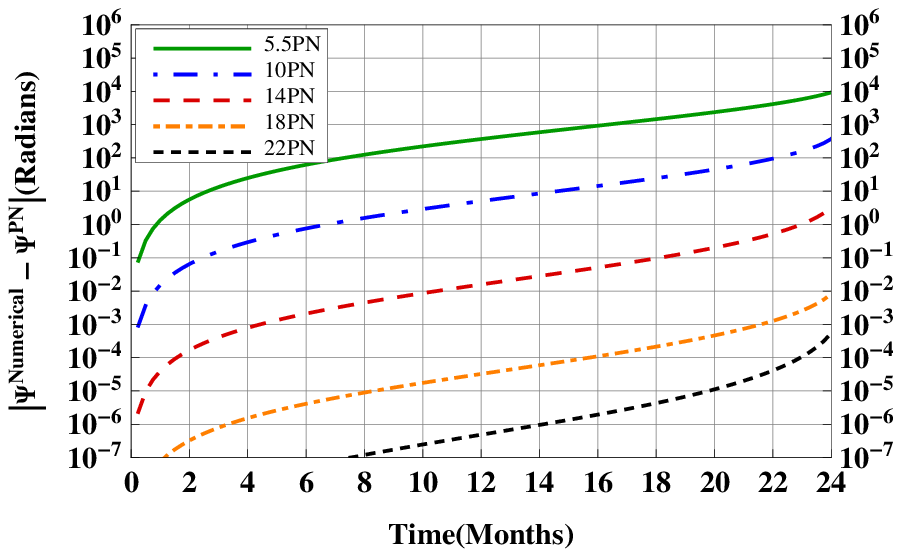}}\quad
		\subfigure[System-I for TaylorEt\label{fig2e}]{\includegraphics[scale=0.85]{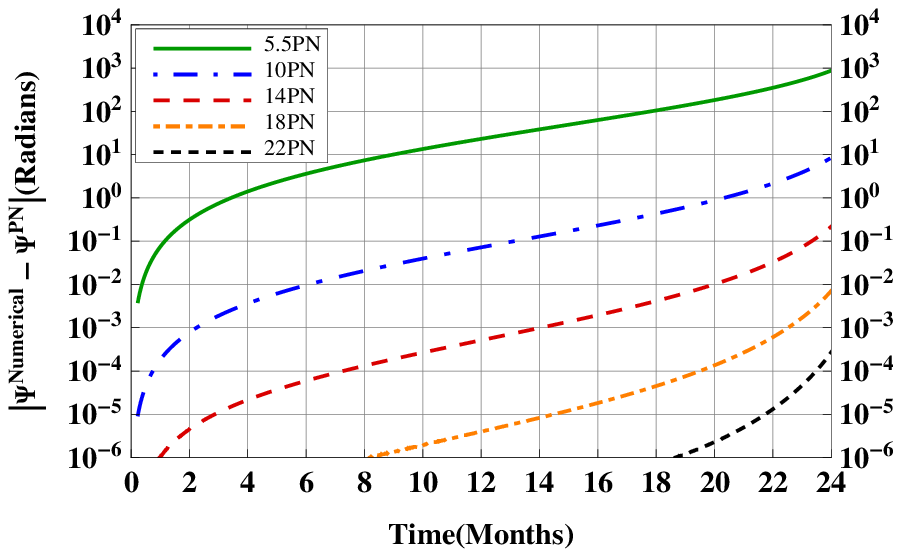}}\quad
    \subfigure[System-II for TaylorEt\label{fig2f}]{\includegraphics[scale=0.85]{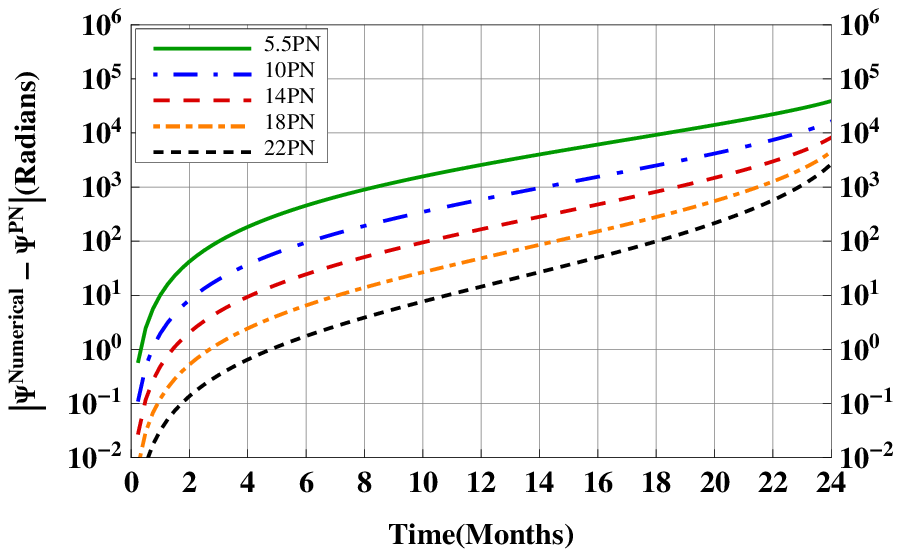}}\quad 
\caption{Absolute value of the dephase between different PN waveforms and numerical waveforms during two-year inspirals as a function of time in months. The left (right) panel shows the dephase for System-I (System-II). Note that the dephase between 18PN (22PN) TaylorT2 waveforms and numerical waveforms at the end of the two-year inspiral for System-I is about $5\times 10^{-9}$ ($10^{-9}$) radians, which falls below the lowest value of dephase shown in Fig. \ref{fig2c}. If the dephase is less than $10$ milliradians, the PN waveforms will provide data analysis accuracies comparable to those provided by the numerical waveforms. Therefore we expect that 18PN waveforms for TaylorT2 are comparable in accuracy of data analysis 
to numerical waveforms for most of the EMRI parameter space of eLISA. 14PN and 18PN waveforms are required for TaylorT4 and TaylorEt respectively, to be comparable to numerical waveforms even in the early inspiral phase (System-I). The accuracy of TaylorT4 and TaylorEt for System-II (which goes up to the LSO) is not comparable to numerical waveform accuracy.  
}\label{fig2}
\end{center}
\end{figure*}

The dephase results are shown in Figs. \ref{fig2a} and \ref{fig2b}. For System-I (II), the absolute values of the dephasing between the TaylorT4 waveforms and the numerical waveforms after the two-year inspiral are about $2\times10^2 (3\times10^4), 7\times10^{-1} (7\times10^3), 6\times10^{-3} (2\times10^3), 5\times10^{-5} (8\times10^2)$ and $5\times10^{-7} (3\times10^2)$ radians for 5.5PN, 10PN, 14PN, 18PN and 22PN respectively. This suggests that 14PN or higher-order TaylorT4 waveforms are comparable in accuracy of data analysis 
to numerical waveforms for the early inspiral phase (System-I) but the accuracy is low for the late inspiral phase (System-II), particularly, near the LSO. 

By comparing with Fig. \ref{fig1}, we see that the dephase for TaylorT4 is a few orders of magnitude worse than that for TaylorT1. This can be explained as follows. In the test particle limit, $dE/dv$, given by 
\begin{equation}
\label{exactenergy}
{dE\over dv} = -v\,\frac{(1 - 6 v^2 )}{(1 - 3 v^2 )^{3/2}},
\end{equation}
goes to zero as one approaches the LSO at $v=1/\sqrt{6}$. Therefore the series expansion of $(dE/dv)^{-1}$ converges very slowly around the LSO. 
Noting that  TaylorT1 and TaylorT4 differ in whether or not the series expansion of $(dE/dv)^{-1}$ is performed in obtaining $dv/dt$, one can expect that TaylorT1 will be more accurate than TaylorT4 near the LSO. 
This suggests  that factorization to avoid a pole at the LSO 
leads  to improvement in the accuracy of $dv/dt$. 
We note that in Refs.~\cite{dis4,din} 
factorization is performed for the energy flux, $\F(v)$ 
to deal with a pole at the light ring, $v=1/\sqrt{3}$. 
Thus, one may also have to factorize the pole at the light ring 
when considering the case beyond the LSO. 


\subsection{Dephase between TaylorT2 and numerical results}
\label{DephaseT2}
TaylorT2 expresses the orbital phase, $\phi(v)$ and the time, $t(v)$ as functions of $v$. For a given time $T$, we solve the equation $T-t(v)=0$ to get the velocity, $v(T)$. Given the velocity one can compute the phase as 
\begin{equation}
\phi(t)=\phi(v(t)).
\end{equation}
$\phi_{\rm ref}$ in Eq.~(\ref{T2a}) is chosen such that $\phi(t=0)=0$.

The dephase results are shown in Figs. \ref{fig2c} and \ref{fig2d}. For System-I (II), the absolute values of the dephasing between the TaylorT2 waveforms and the numerical waveforms after the two-year inspiral are about $9\times10^1 (10^4), 4\times10^{-2} (6\times10^2), 5\times10^{-6} (5), 5\times10^{-9} (10^{-2})$ and $10^{-9} (8\times10^{-4})$ radians for 5.5PN, 10PN, 14PN, 18PN and 22PN respectively. Therefore for TaylorT2, 10PN (18PN) waveforms may provide accuracies comparable to those provided by numerical waveforms for System-I (System-II). 

As can be seen by comparing with Fig. \ref{fig1}, the dephase of TaylorT2 is comparable or lesser than TaylorT1 during the inspirals. However, one needs to keep in mind that for calculating the phase from the TaylorT2 approximant, a pair of transcendental equations needs to be solved, which can be very time consuming and expensive.                    

\subsection{Dephase between TaylorEt and numerical results}
\label{DephaseEt}
TaylorEt expresses $d\phi/dt$ and $d\zeta/dt$ as power-series expansions of $\zeta=-2E/\nu$. For a given time $T$, we solve the equation 
$T-\int_{\zeta_0}^{\zeta(T)} 1/(d\zeta/dt) \,d\zeta=0$ to get $\zeta(T)$, where $\zeta_0$ and $\zeta(T)$ are the values of $\zeta$ at $t=0$ and time $T$, respectively.

The phase can now be evaluated as
\begin{equation}
\phi(t)=\int_{0}^{t} (d\phi/dt') \,dt' =\int_{\zeta_0}^{\zeta(t)} 
\frac{(d\phi/dt')}{(d\zeta/dt')} \,d\zeta, 
\end{equation}
where $(d\phi/dt')$ and $(d\zeta/dt')$ are given in Eq.~(\ref{Et}). 

The dephase results are shown in Figs. \ref{fig2e} and \ref{fig2f}. For System-I (II), the absolute values of the dephasing between the TaylorEt waveforms and the numerical waveforms after the two-year inspiral are about $10^3 (6\times10^4), 9 (2\times10^4), 3\times10^{-1} (9\times10^3), 8\times10^{-3} (7\times10^3)$ and $4\times10^{-4} (4\times10^3)$ radians for 5.5PN, 10PN, 14PN, 18PN and 22PN respectively. This suggests that 18PN and 22PN TaylorEt waveforms are comparable in accuracy of data analysis 
to numerical waveforms for the early inspiral phase (System-I) of eLISA frequency band but the accuracy is low for the late inspiral phase (System-II).

As in the case of TaylorT4 in Sec.~\ref{DephaseT4}, 
the reason that the performance of TaylorEt is much worse than that of 
TaylorT1 can be related to the poor convergence of the series expansion 
of $(dE/dv)^{-1}$ or $dv/dt$ around the LSO. 
Solving $\zeta=-2E/\nu$ iteratively,\footnote{One can solve $\zeta=-2E/\nu$ 
to obtain the explicit expression for $v(\zeta)$ 
without performing a series expansion in terms of $v$. 
But the explicit expression for $v(\zeta)$ may not be useful to implement 
since it contains the square root of polynomial functions of $\zeta$.
} 
the new variable $\zeta$ in TaylorEt 
can be related to $v$ as in Eq.~(\ref{x}). 
One can also derive the same relation using 
$v(\zeta)=\int (d\zeta'/dv)^{-1}\,d\zeta'$. 
Noting $(d\zeta/dv)^{-1}=-\nu\,(dE/dv)^{-1}/2$, we see that
the integrand of $v(\zeta)$ contains a pole at the LSO. 
Then, one may expect that $v$ as a series expansion in terms of $\zeta$ 
does not converge very well around the LSO. 
Hence functions of $\zeta$ in TaylorEt, which are computed by using 
a series expansion of $v(\zeta)$, will not converge well. 


\subsection{Dephase between TaylorT3 and numerical results}
\label{DephaseT3}
TaylorT3 gives the orbital phase, $\phi(t)$, and the instantaneous gravitational wave frequency, $F(t)$, as functions of $\theta(t)$, which is a function of time. For any given time $t$, we can find the phase as
\begin{equation}
\phi(t)=\phi(\theta(t)). 
\end{equation}

As in TaylorT2, $\phi_{\rm ref}$ is chosen such that $\phi(t=0)=0$.

\begin{figure}[htb]
\begin{center}
 
\includegraphics[scale=0.75]{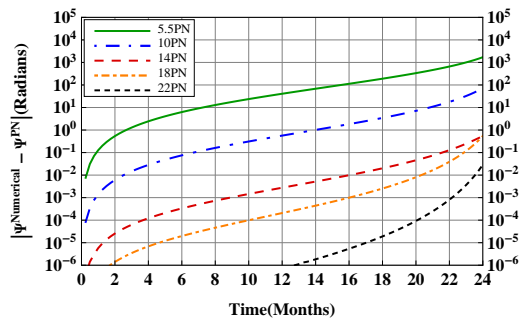}
		  
\caption{Absolute value of the dephase between TaylorT3 PN waveforms and numerical waveforms during two-year inspirals for System-I as a function of time in months. TaylorT3 is found to behave poorly for System-II as it goes up to the last stable orbit (LSO). We expect that 22PN TaylorT3 waveforms are required to get data analysis accuracies comparable to those provided by numerical waveforms for System-I.  
}\label{fig3}
\end{center}
\end{figure}

The dephase results for System-I are shown in Fig. \ref{fig3}. For System-I, the absolute values of the dephasing between the TaylorT3 waveforms and the numerical waveforms after the two-year inspiral are about $2\times10^3, 9\times 10^1, 7\times 10^{-1}, 7\times 10^{-1}$ and $3\times 10^{-2}$ radians for 5.5PN, 10PN, 14PN, 18PN and 22PN respectively. Therefore we see that 22PN TaylorT3 waveform is required to get data analysis accuracies comparable to numerical waveform accuracy for System-I. 

The TaylorT3 approximant is not accurate in the case of System-II as it goes up to the LSO ($v=1/\sqrt{6}$). 
Even for System-I we find the dephase is a few orders of magnitude higher than that for TaylorT1. 
One also finds that the value of $F^{(\rm T3)}(t)$ becomes very large 
for $\theta(t)\ge 0.67$, and for System-II one cannot find a 
$t_{\rm ref}$ consistent with the one derived for TaylorT2. 
The reason for this is again similar to the reason for the poor behavior of TaylorT4 as compared to TaylorT1. In the TaylorT3 approximation, one derives $v(t)$ as a series of $t$, 
i.e. $\theta$, by iteratively inverting $t(v)$ in TaylorT2. 
$v(t)$ as a series of $t$ can also be derived using 
$v(t)=\int (dv/dt')\,dt'$. 
Since the integrand of $v(t)$, $dv/dt$, has a pole at LSO, 
one will expect that $v(t)$ as a series of $t$ does not converge well 
around LSO. Thus, functions of $\theta$ in TaylorT3, computed 
using $v(t)$ in a series of $t$, will not converge well. 

\section{Conclusions}
\label{conc}
Using the 22PN expression for flux, $\F(v)$, derived in Ref. \cite{f22}, we calculated the TaylorT1, TaylorT2, TaylorT3, TaylorT4 and  TaylorEt 
approximants up to 22PN order for a test particle in a circular orbit around a Schwarzschild black hole. We evaluated the performance of the PN waveforms by calculating the gravitational wave phase predicted for two EMRI systems, System-I ($m_1/m_2 = 10^{-4}$) and System-II ($m_1/m_2 = 10^{-5}$) during two-year inspirals. System-I (System-II) corresponds to the early (late) inspiral phase of the eLISA frequency band. The phase predicted by PN waveforms is compared with the phase predicted by waveforms resulting from high-precision numerical solutions of the Teukolsky equation 
for the same inspirals. For accurate eLISA EMRI parameter estimation with these PN waveforms, we need the difference of the phases, the dephase, to be less than $10^{-2}$ radians~\cite{accu}.

We found that the dephase between the 22PN waveforms and numerical waveforms after a two-year inspiral for System-II is smaller than $10^{-2}$ and $10^{-3}$ 
radians for TaylorT1 and TaylorT2 respectively. Therefore we expect that these 22PN waveforms can be used to attain data analysis accuracies comparable to 
those provided by high-precision numerical waveforms for most of the parameter space of EMRIs. 
Moreover, for the early inspiral phase, 10PN waveforms for 
TaylorT1 and TaylorT2 may be used for data analysis. 

However, the dephase of TaylorT4, TaylorEt and TaylorT3 waveforms goes to 
values higher than $10^{2}$ radians for System-II. This suggests that these 
approximants cannot be used for data analysis of late inspirals. 
We note that our results reinforce investigations in Ref.~\cite{biops} 
that TaylorEt and TaylorT3 are recommended not to be used 
for data analysis of comparable mass binaries.

For System-I we found that 14PN or higher PN order waveforms are required 
for TaylorT4, TaylorEt and TaylorT3 to achieve 
comparable results in data analysis to using high-precision numerical 
waveforms even in the early inspiral phase.  
We also found that the reason the dephases of TaylorT4, 
TaylorEt and TaylorT3 waveforms 
are much larger than those of TaylorT1 and TaylorT2 
may be related to the fact that $(dE/dv)^{-1}$ has a pole at the LSO. 
This suggests that when constructing templates for coalescing compact binaries, 
approximants avoiding the pole at the LSO arising from the energy function
by factorizing it as in TaylorT1, or those 
introducing a new variable which cancels the pole may perform better.
We hope that these studies also provide insights to construct 
more efficient templates for coalescing compact binaries in the 
comparable mass case. 
Lastly, the analytical expressions for the various approximants could also
be useful for studies related to the ground-based detectors. 

\begin{center}
\Large{\bf Acknowledgments}
\end{center}
V.V. and A.C. thank the Raman Research Institute for support during the Visiting Student Program during which this work was initiated. R.F. is grateful for the support of the European Union FEDER funds, 
the Spanish Ministry of Economy and Competitiveness (Projects No. 
FPA2010-16495 and No. CSD2007-00042) and the Conselleria 
d'Economia Hisenda i Innovacio of the Govern de les Illes Balears. 
\bibliographystyle{apsrev4-1}
\bibliography{reference}
\end{document}